\documentclass[12pt] {article}

\usepackage{epsfig}
\usepackage{subfigure}

\def\e{\begin{equation}}
\def\f{\end{equation}}
\def\%#1{\mbox{\boldmath $#1$}}
\def\=#1{\overline{\overline #1}}

\def\_#1{{\bf #1}}
\def\o{\omega}
\def\E{\epsilon}

\def\.{\cdot}

\def\##1{{\bf#1\mit}}

\def\am{\left(\begin{array}{c}}
\def\amm{\left(\begin{array}{cc}}
\def\a{\end{array}\right)}
\def\v{\vspace{0.5cm}}

\title{Light-Weight Base Station Antenna with Artificial Wire Medium Lens}

\author{Pekka Ikonen$^1$, Mikko K\"arkk\"ainen$^1$, Constantin Simovski$^{1,2}$, Pavel Belov$^{2}$, and\\Sergei Tretyakov$^1$}

\date{$^1$Radio Laboratory/SMARAD, Helsinki University of
Technology\\P.O. Box 3000, FI-02015 TKK, Finland\\ \v $^2$Dept.~Physics, State University of Informational Technology,\\
Mechanics and Optics at St.~Petersburg,\\ Sablinskaya 14, 197101,
St. Petersburg, Russia}

\begin{document}

\maketitle

{\center \large

\v

Address for correspondence:

Pekka Ikonen, \\ Radio Laboratory, Helsinki University of
Technology,\\ P.O. Box 3000, FI-02015 TKK, Finland.

Fax: +358-9-451-2152

E-mail: pikonen@cc.hut.fi

}

\parindent 0pt
\parskip 7pt

\vspace{0.5cm}

\baselineskip = 1.3em

\begin{center}
\section*{Abstract}
\end{center}
In this paper we study the possibility of utilizing a loaded wire
medium as an artificial material for beam shaping element in base
station antenna applications. The wires in the lattice are loaded
with a proper capacitive impedance to come up with a structure
effectively operating as a dielectric lens at the UMTS (Universal
Mobile Telecommunications System) frequency range. It is proven with
full-wave simulations that the interpretation of a finite size wire
lattice as a dielectric lens is physically sound even with a
moderate number of lattice periods. A prototype having a
mechanically reconfigurable beam width is constructed and measured.
It is shown that very promising performance can be achieved with a
rather simple structure and a cheap manufacturing process.

\textbf{Key words}: Wire medium, artificial dielectrics, lens
antenna

\newpage

\section{Introduction}

It is expected that the enormous growth of the traffic load in the
wireless networks will continue in the near future. The transition
era from the 3$^{\rm rd}$ generation systems to 4G networks leads to
applications that will require huge data rates and therefore
extremely efficient network planning \cite{Webb,Mishra}.
Undoubtedly, novel implementation techniques are needed to  lighten
the pressure concerning the improvement of base station antennas.
Ultimately, the antenna must be robust and cheap for mass production
implying that the structure should be small and simple. However,
sectorized cells and the natural evolutional stages of the network
often set the demand for the antenna to operate in different modes
with different service areas. Therefore, the antenna should offer a
possibility for a tunable beam width or a switchable beam direction.
Incorporating intelligence in a miniaturized and simple antenna
structure while still maintaining the efficiency high, is not an
easy task.

The operational principle of a medium consisting of  periodically
arranged wires has been known for a long time \cite{Brown,Rotman}.
This kind of \emph{wire medium} (also called \emph{rodded medium})
is known to operate as an artificial dielectric with negative
effective permittivity at low frequencies. Due to the periodic
nature of the wire medium, the medium can be considered as an
electromagnetic band-gap (EBG) structure introducing frequency bands
in which electromagnetic waves cannot propagate \cite{Joannopoulos}.
A detailed analysis of the electromagnetic band structures of the
wire medium can be found e.g. in \cite{Kuzmiak,Nicorovici}. The
transmission and absorption properties of a two-dimensional (2D)
lattice of conducting cylinders have been presented \cite{Sigalas},
and the effective electronic response of a system of metallic
cylinders has been introduced \cite{Pitarke}.

The nature of wave propagation inside the wire medium  and its
dispersion properties have also been subjects of comprehensive
studies. In \cite{Moses-Engheta} the wave propagation inside the
wire medium was studied in detail.
Comprehensive analytical study of the dispersion and reflection
properties of the wire medium can be found e.g. in
\cite{Belov-JEWA,Belov-PhysE,Belov-PhysB}. Authors of
\cite{Belov-PhysE} considered also a wire medium periodically loaded
with bulk reactances and presented the dispersion relation for this
kind of \emph{loaded wire medium}. A quasi-static  model for the
loaded wire medium can be found in \cite{Maslovski}.

Recently, extensive research has been devoted to the  utilization of
the wire medium in microwave applications. The transmission
properties of loaded and unloaded wire medium have been
theoretically and experimentally analyzed in \cite{Lourtioz}. In
\cite{Lustrac1} the authors presented an EBG structure consisting of
cut metal strips mounted with pin-diodes. Depending on the diode
biasing the structure was shown to prohibit the propagation of the
electromagnetic waves or to create a very directed beam for a planar
antenna. Discussion about the performance of antennas inside wire
lattices has been presented in \cite{Poilasne_the}, and authors of
\cite{Poilasne2} presented an experimental demonstration for the
radiation characteristics of a conventional dipole inside the wire
medium. The utilization of dipole arrays (somewhat resembling the
case of cut wires with the introduced cap width extending towards
infinity) as a beam shaping lens was introduced in the beginning of
80's by Milne \cite{Milne}. A mathematical design procedure for a
non-homogenous wire medium lens antenna can be found in
\cite{Silveirinha2}. Analytical and numerical results for the
radiation properties of a simple radiator in the vicinity of the
loaded wire medium were presented in \cite{Simovski1}, and an
experimental demonstration of the feasibility to use the loaded wire
medium with compact and directive antenna applications can be found
in \cite{Ikonen}.

\begin{figure}[t!]
\begin{center}
\includegraphics[width=10cm]{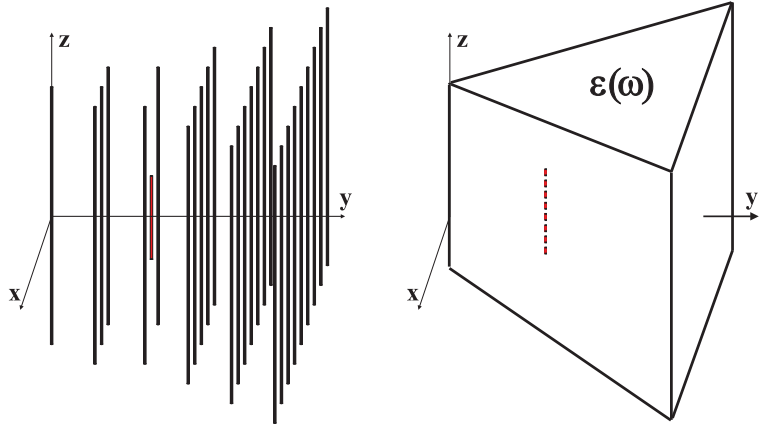}
\end{center}
\caption{Loaded wire medium constituting an artificial dielectric
lens.} \label{lens}
\end{figure}

In the present paper we utilize the properties of the loaded wire
medium \cite{Belov-PhysE,Simovski1} to design a compact dual-mode
base station antenna having a mechanically reconfigurable beam
width. After capacitive loading the plasmon-like band-gap inherent
to the conventional wire medium disappears and the structure becomes
transparent for radiation \cite{Belov-PhysE,Lourtioz}. Moreover, the
proper loading allows us to reduce dramatically the size of the
radiator without a loss of the radiation efficiency
\cite{Simovski1}. It is shown below that the interpretation of the
loaded wire medium as an artificial dielectric lens is possible even
with a moderate number of lattice periods. Around the frequency
range of interest the wire lattice operates as a finite size block
of artificial dielectric as schematically depicted in
Fig.~\ref{lens}. The key benefits of the proposed antenna structure
are low cost, cheap manufacturing process, and extremely light
weight. An antenna prototype is manufactured and measured. It is
shown that the theory of an aperture radiator is applicable in the
designing process of the wire medium lens antenna.

\section{Loaded wire medium: Revision of the dispersion properties}

\begin{figure}[t!]
\begin{center}
\includegraphics[width=8.0cm]{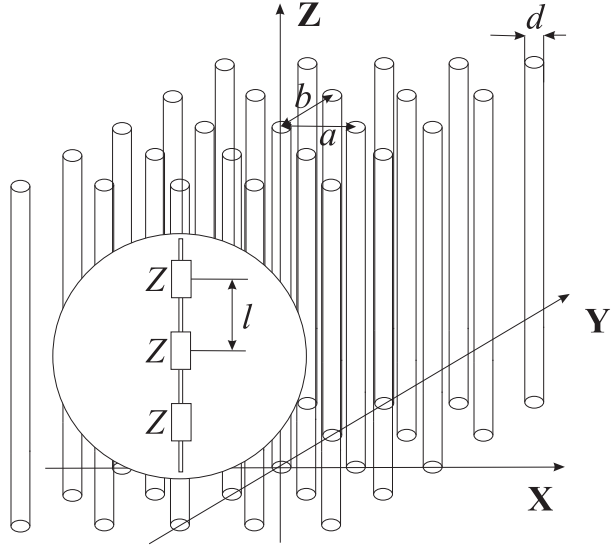}
\end{center}
\caption{Loaded wire medium.} \label{loaded_wire_medium}
\end{figure}

A schematic illustration of the loaded wire medium is shown in
Fig.~\ref{loaded_wire_medium}. It is known that the conventional
wire medium (solid wires) exhibits a plasmon-like band-gap starting
from zero frequency \cite{Nicorovici,Belov-PhysE,Lourtioz}. This
means that at low frequencies electromagnetic waves cannot propagate
through the wire medium (except to the direction along the wires).
This further implies that if a radiator is brought inside
conventional wire medium, a channel must be configured in the
structure to permit the emission of electromagnetic waves (see e.g.
\cite{Poilasne2} for experimental demonstration). In addition to
this, the transversal size of such a structure is necessarily rather
large due to destructive interaction between the source current and
the currents flowing in the closely located solid wires. When the
wires are loaded with bulk capacitances the low  frequency band-gap
disappears \cite{Belov-PhysE,Lourtioz}. Moreover, it was
analytically shown in \cite{Simovski1}, and experimentally proven in
\cite{Ikonen} that the problem of field cancelation can also be
overcome with the use of bulk capacitances.

\begin{figure}[t!]
\begin{center}
\includegraphics[width=10cm]{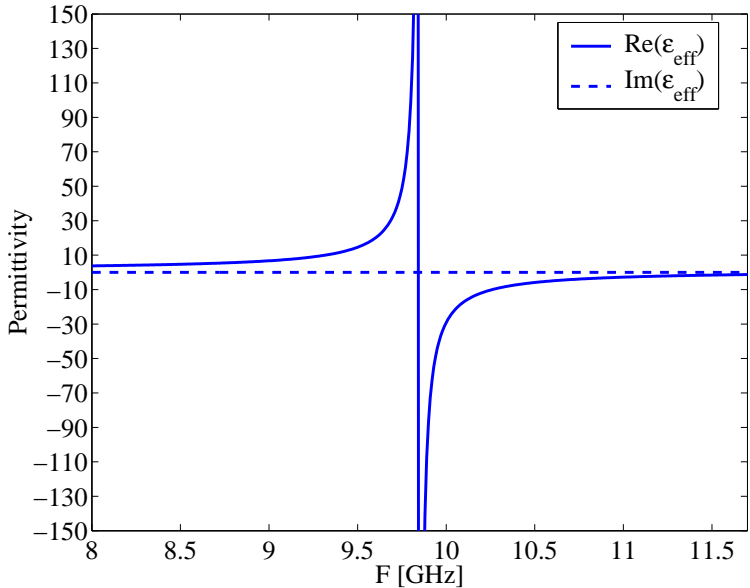}
\end{center}
\caption{Dispersive behavior for $\E_{\rm eff}$ of the loaded wire
medium.} \label{eps_ex}
\end{figure}

Let us consider the loaded wire medium periodically loaded with bulk
capacitances as shown in Fig.~\ref{loaded_wire_medium} ($Z$ =
$-j|Z|$). In \cite{Belov-PhysE} the dispersion equation for such a
medium was formulated in terms of Floquet harmonics, scattered by
the set of parallel wire grids periodically located along the
$x-$axis (with period $a$, see Fig.~\ref{loaded_wire_medium}).
Authors of \cite{Belov-PhysE} presented also a simplified expression
for the full dispersion equation applicable for dense wire lattices
(the lattice periods $a$ and $b$ are much smaller than the
wavelength in the matrix material). Moreover, special attention was
devoted to waves traveling orthogonally to the wire axis (the
$z$-component of the propagation factor is zero, $q_{\rm z}$ = 0).
This special case was considered at low frequencies compared to the
first lattice resonance of the loaded wire medium. At these
frequencies the magnetic properties of the medium are negligible and
the effective medium model becomes valid. With the aforementioned
assumptions, the simplified dispersion equation for dense wire
lattices was reformulated using the resonant effective permittivity
\e q_{\rm x}^2 + q_{\rm y}^2 = \omega^2\epsilon_{\rm
eff}(\omega)\E\mu = k^2{\epsilon_{\rm eff}(\omega)}, \label{ISO1} \f
where $\E$ and $\mu$ are the material parameters of the matrix
material, and $k$ is the wave number in the matrix material. For a
capacitively loaded wire medium the resonant effective permittivity
takes the form \cite{Belov-PhysE} \e \E_{\rm eff}(\o) = 1 +
\frac{C/(\E_0s^2)}{1 - \o^2/\o_0^2}, \label{permittivity} \f where
$C$ is the load capacitance $C_0$ multiplied by the period of
insertions $l$, and $s$ is a geometrical parameter defined below.
The resonant frequency can be expressed as \cite{Belov-PhysE} \e
\o_0^2 = \frac{2\pi/(\mu_oC)}{\ln{\frac{s}{\pi{d}}} + F(r)}, \quad s
= \sqrt{ab}, \quad r = a/b, \label{omega_0} \f where $a$ and $b$ are
the lattice periods, and $d$ is the diameter of the wires. $F(r)$ is
defined as \e F(r) = -\frac{1}{2}\log{r} + \sum_{n =
1}^{+\infty}\bigg{(}\frac{\coth(\pi{n}r)-1}{n}\bigg{)} +
\frac{\pi{r}}{6}. \label{Fr} \f The real part of the effective
permittivity of a wire medium effectively loaded with a certain
capacitance per unit length is shown in Fig.~\ref{eps_ex}. We can
see from Fig.~\ref{eps_ex} that the permittivity obeys the Lorenzian
type dispersion rule. Before the first stop band ($F$ $<$ 9.85 GHz)
the refractive index is a monotonically growing function and the
loaded wire medium operates as a usual artificial dielectric. At
frequencies  9.85 $\leq F <$ 13.7  GHz the medium can be considered
as an artificial plasma (Re$\{\epsilon_{\rm eff}\}$ $<$ 0).

{

Relatively far away from the structure resonance the effective
permittivity has nearly a constant value. For reference, around 2
GHz the real part of the effective permittivity equals approximately
2.0. Rather low and frequency-independent value for the permittivity
of the lens can be considered as an advantage for the lens design.
The aforementioned supports the idea of utilizing the loaded wire
medium as a light-weight artificial dielectric lens, as
schematically depicted in Fig.~\ref{lens}.

Note that this principle of utilization is fundamentally different
from the angular filtering techniques \cite{Tayeb} introduced in the
literature for different types of EBG structures
\cite{Lourtioz,Lustrac1,Cheype,Temelkuran}. The known beam shaping
methods utilize defects in the EBG structure to create a
transmission peak inside the original stopband. It is known
\cite{Tayeb} that a slight local change in the crystal period leads
to localized resonant modes, and can be used for the realization of
devices radiating energy in a very narrow angular range. At
frequencies lying inside the original band-gap of the crystal,
discontinuous wires \cite{Lustrac1} or dielectric rods
\cite{Cheype,Temelkuran} located in front of a low gain radiator can
also be used as simple angular filters. One of the biggest
disadvantages of this utilization technique  is a very narrow
operational bandwidth. We, on the contrary, propose the utilization
of the EBG structure far below the first stopband and regard the
loaded wire medium as a piece of a usual homogenous dielectric.

It is also worth noticing that due to very high signal power levels
used in base stations (e.g. Kathrein \cite{Kathrein} reports the
maximum power per input to be 300 W for most of the products),
active components cannot be used in the lens due to unavoidable
intermodulation products leading to distorted frequency spectrum.
Therefore we seek for a solution where the bulk capacitances are
implemented without using active components.

\section{Designing the novel antenna structure}

\subsection{Desired characteristics}

The goal of the following antenna design procedure is to come up
with a compact base station antenna structure having two modes
corresponding to two beam widths in the H-plane (the plane
orthogonal to the wires). The carrying idea is to design as simple
and robust antenna as possible. The beam width is to be reconfigured
mechanically to maintain the simplicity of the design, and to avoid
problems caused by active components when using high signal power
levels. The targeted radiation characteristics are presented below:
\begin{enumerate}
\item Frequency range 1920 -- 2170 MHz (UMTS FDD range)
\item Maximum transversal size $<$ 15 cm\footnote{The size of the reflector excluded.}
\item Backward radiation as small as possible
\item Half-power beam width in H-plane 65$^{\circ}$ / 85$^{\circ}$
\end{enumerate}
Other important features for a design close to an  end product are
the E-plane beamwidth, E-plane side lobe level, the gain, and the
longitudinal size of the structure. The main practical importance of
the present work is the demonstration of the beam switching with the
help of the artificial wire medium lens. However, to be assure for
the feasibility of the structure in base station antenna
applications, we will also introduce the radiation pattern in
E-plane, and the simulated gain values. The optimization of the
antenna matching, and the details in the actual mechanical switching
procedure are not discussed in the present work.

The main parameters characterizing the feasibility of the antenna
structure are the deviation in the input impedance, front-to-back
ratio (FBR), robustness and the H-plane beamwidth. Ideally, the only
changing parameter between the modes should be the H-plane
beamwidth. The robustness refers to the fact that the radiation
characteristics of the antenna should not change dramatically when
the parameters of the structure (e.g. the lattice parameters $a,b$,
the value of the load capacitance $C_0$) are changed slightly. This
is very important in the view of the manufacturing process. To
maintain the design as simple as possible, we seek for a solution
where the structure consists of two parts. Namely, the antenna has a
fixed part and a removable part. The fixed part consists of the
radiator accompanied with a possible reflector, whereas the
removable part is formed by the wires located in front of the
radiator. Fixing the radiator and the reflector sets an additional
demand for the removable part. In addition to producing different
beam widths, the removable part should not change strongly the input
impedance seen at the antenna terminals. This ensures a reasonable
matching level with both modes.

\subsection{Numerical design}

\begin{figure*}[b!]
\centerline{\subfigure[Wide lens topology.]
{\includegraphics[width=5.5cm]{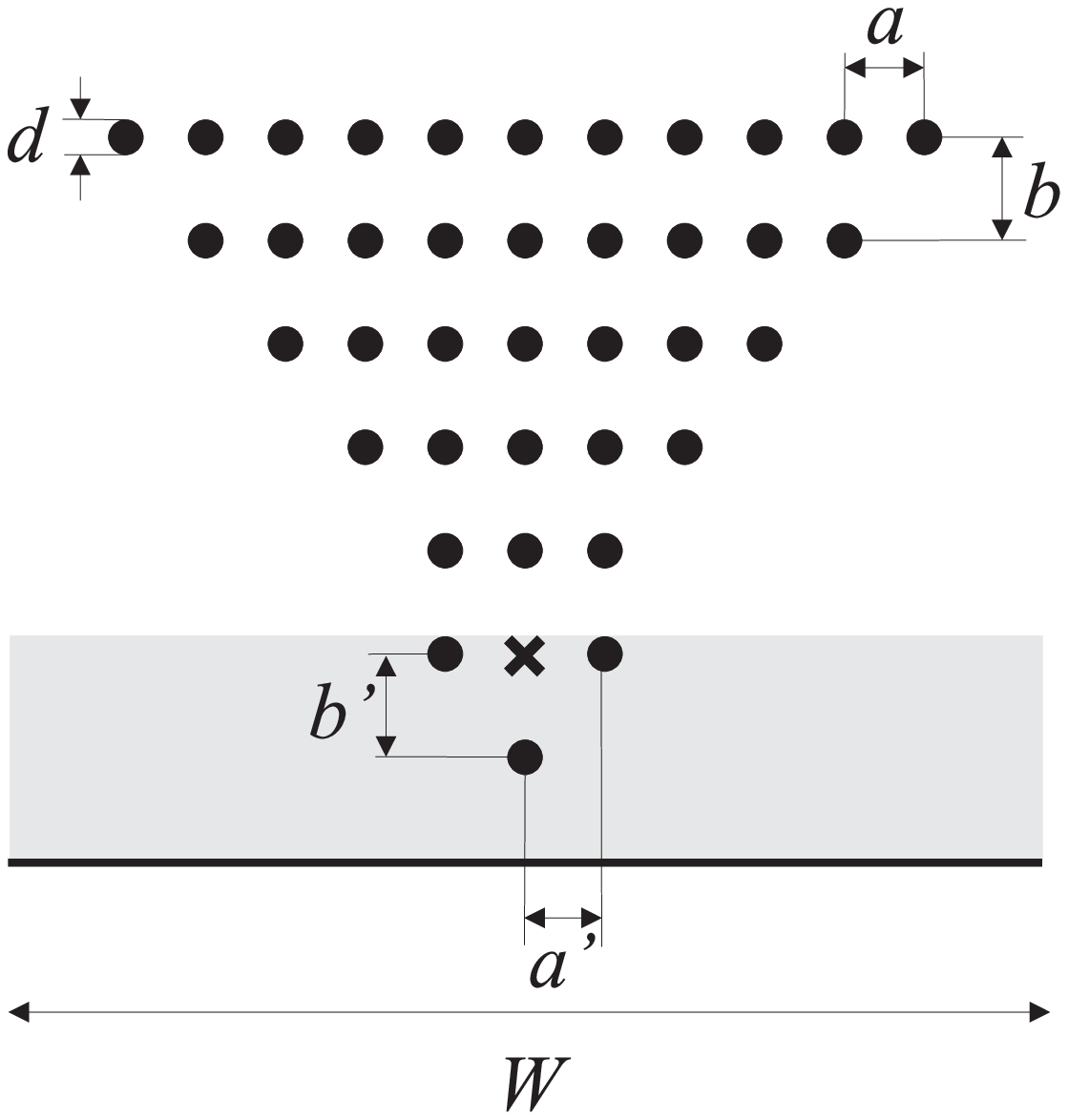}
\label{fig_first_case}} \hfil \subfigure[Narrow lens
topology.]{\includegraphics[width=5.5cm]{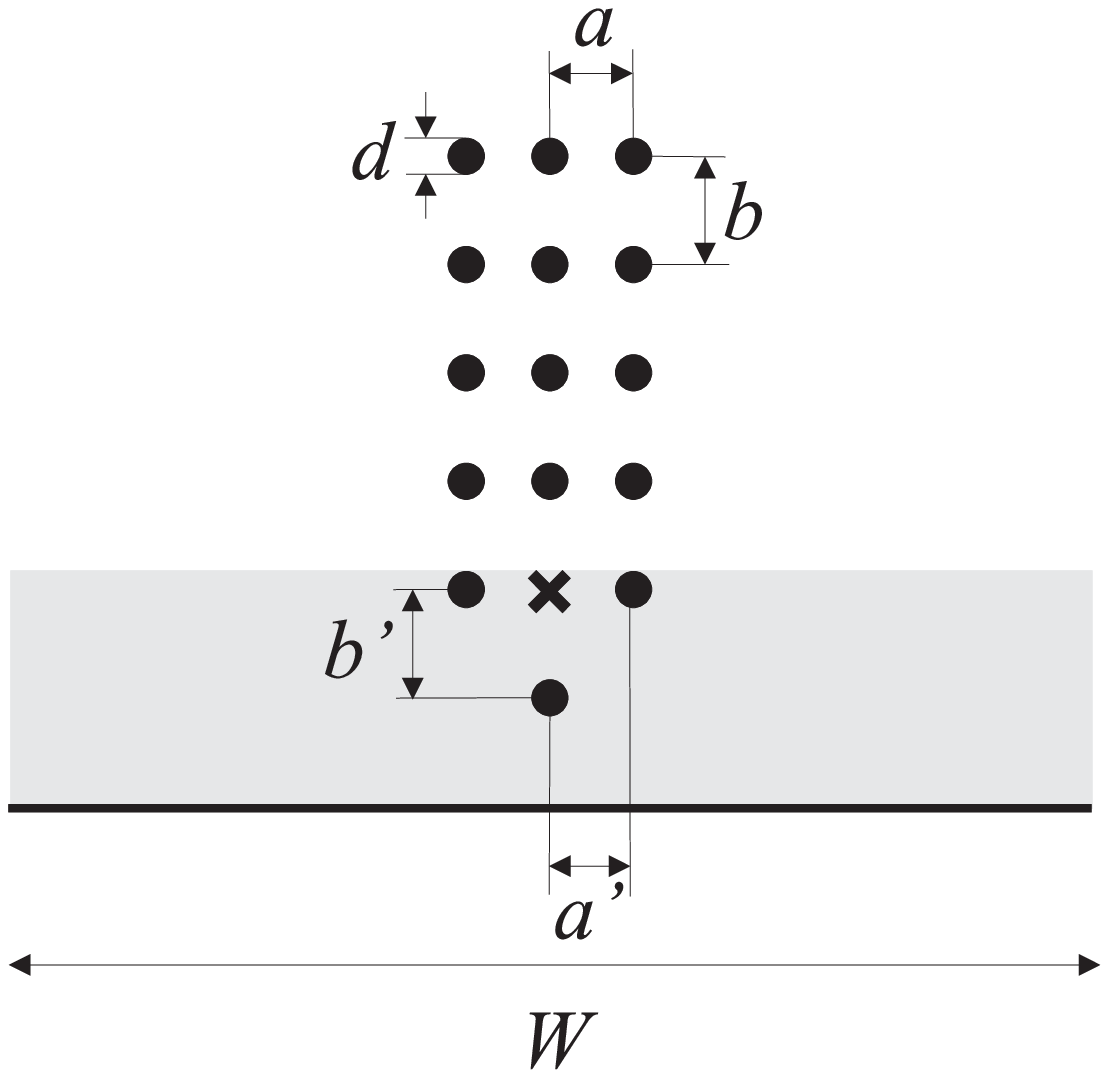}
\label{fig_second_case}}} \caption{The transversal geometry of the
prototype (the fixed part is shown shadowed). The black circles
denote capacitively loaded wires, the cross denotes the dipole. The
horizontal line denotes the reflector. $a$ = 11.5 mm, $b$ = 12.5 mm
(wide lens), $b$ = 10.0 mm (narrow lens), $a'$ = 11.5 mm, $b'$ =
12.5 mm, $d$ = 1.125 mm. The insertion period of the loads $l$ =
5.55 mm, total length of the wires $L$ = 200 mm. $W$ = 147.2 mm. The
value of the load capacitance in the fixed part, the wide lens, and
the narrow lens respectively: $C_{\rm f}=$ 0.125 pF, $C_{\rm w}=$
0.125 pF, $C_{\rm n}=$ 0.150 pF.} \label{proto_geom}
\end{figure*}

Suitable geometry for the prototype leading to the desired
performance was found with numerical simulations conducted with a
commercial method of moments (MoM) simulator \verb"FEKO"
\cite{Feko}. It was suggested in \cite{Simovski1} and experimentally
proven in \cite{Ikonen} that to achieve a zenithal radiation pattern
in the H-plane, the optimal number of loaded wires with a
half-wavelength dipole is approximately 20--30 and the shape of the
lattice is close to a triangular one. This estimation laid the basis
for the optimization process.

Fig. \ref{proto_geom} shows the final outcome of  the designing
process. As we can see there is a considerable difference between
the topology of the wide and the narrow lens. This is, however, well
predictable according to the theory of an aperture radiator
\cite{Balanis}: As the far field radiation pattern of an aperture
radiator (e.g. a lens illuminated with a low gain radiator) is the
Fourier transform of the illuminating field distribution, the wider
is the aperture, the narrower is the beam. To suppress the backward
radiation a metal reflector is used. The reflector can also be made
of simple solid wires (with a distance between the wires much
smaller than $\lambda$) to lighten the wind load. The transversal
size with the wide lens (excluding the reflector size) is 11.5
$\times$ 8.75 cm$^2$ (0.77 $\times$ 0.60 $\lambda^2$) implying that
the structure is very compact. The transversal size of the structure
with the narrow lens is 2.3 $\times$ 8.4 cm$^2$ (0.15 $\times$ 0.55
$\lambda^2$). The distance between the dipole and the closest wire
is $a'$ = 11.5 mm $\sim$ $\lambda$/13.

In the present design the beam switching (reconfiguration) function
is implemented by manually changing the removable part in front of
the radiator. Other possibilities (not studied in the present paper)
could include fixing the lens topology and modifying the impedance
loading by mechanically or electronically tuning the load
capacitances. This, however, adds complexity to the design, and
might require the use of active components in the lens.

Fig. \ref{fig_sim} introduces the simulated far field patterns in
the H- and E-planes. The main simulated parameters are gathered in
Table 1. Presented simulations have been conducted at the UMTS FDD
uplink (UL) and downlink (DL) center frequencies, namely at $F_{\rm
UL}=$ 1.95 GHz and $F_{\rm DL}=$ 2.14 GHz. As can be seen, the beam
widths in the H-plane are very close to the goal values 65$^{\circ}$
and 85$^{\circ}$ with the wide and narrow lens respectively. The
beam width in the E-plane is typical when using only one dipole as a
source. The simulated gain values (at the center frequency $F_{\rm
c}$= 2.045 GHz) with the wide and narrow lens are 10.9 and 9.6 dBi,
respectively. A design closer to an end product could have e.g. a
collinear array of crossed dipoles and a total height close to one
meter. This would significantly narrow the E-plane beam width and
increase the gain of the structure.

\begin{figure*}[t!]
\centerline{\subfigure[The simulated H-plane pattern.]
{\includegraphics[width=7.5cm]{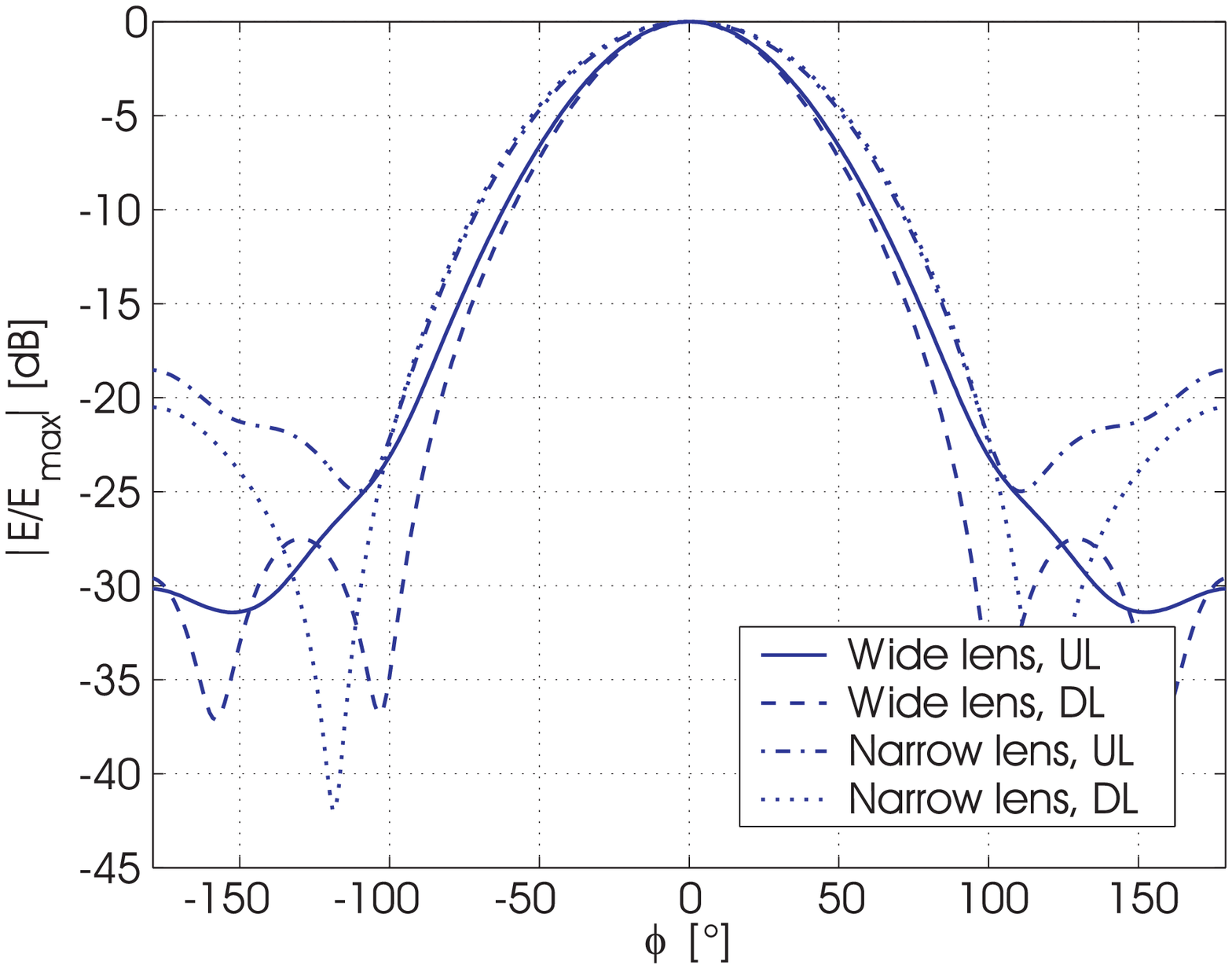} \label{hp}} \hfil
\subfigure[The simulated E-plane
pattern.]{\includegraphics[width=7.5cm]{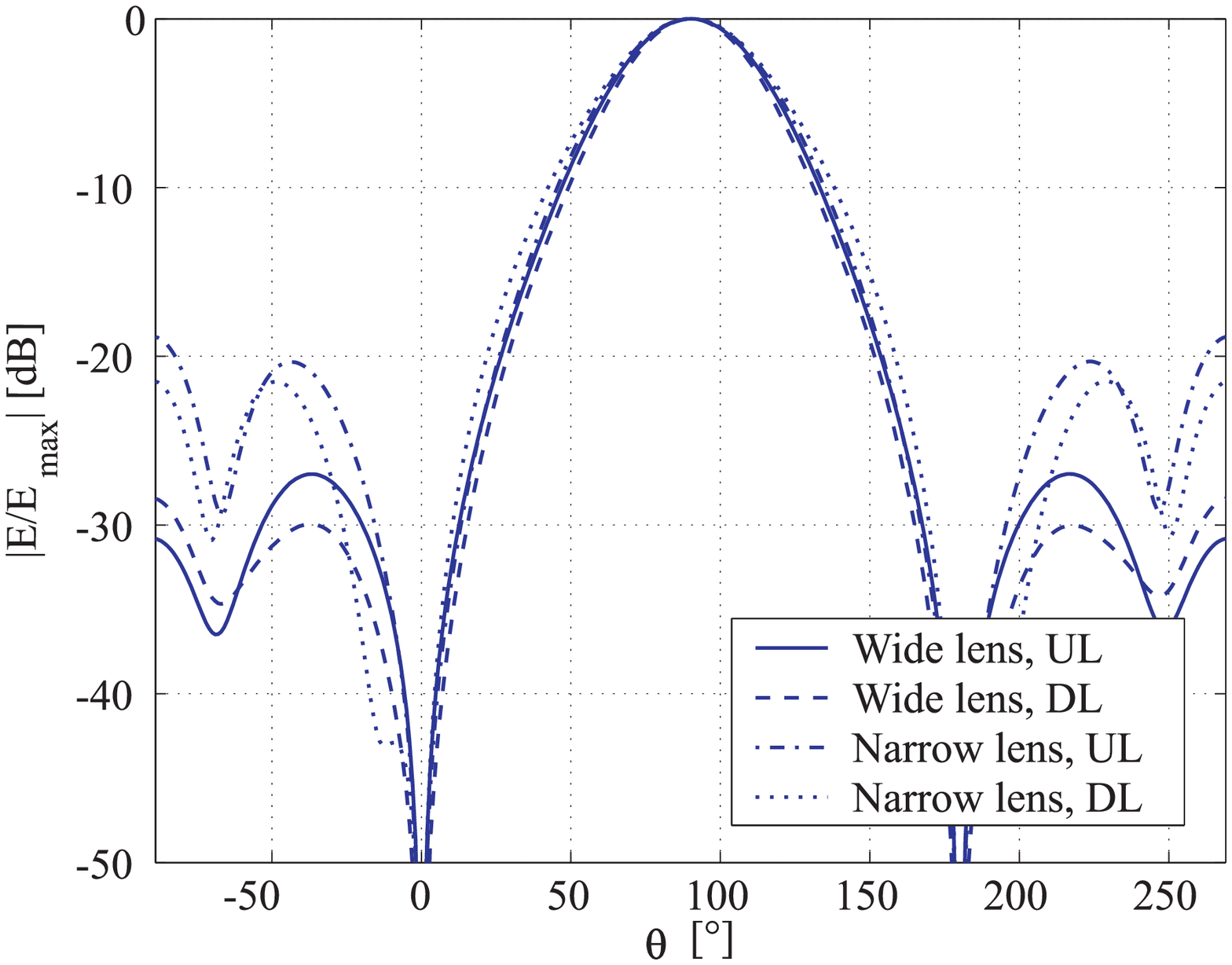}
\label{ep}}} \caption{The simulated radiation patterns. ``UL''
corresponds to $F = F_{\rm UL} = 1.95$ GHz, ``DL'' corresponds to $F
= F_{\rm DL} = 2.14$ GHz.} \label{fig_sim}
\end{figure*}

\begin{table}[b!]
\renewcommand{\arraystretch}{1.3}
\label{protosim} \centering \caption{The simulated main radiation
characteristics for  the prototype.}
\begin{tabular}{|c|c|c||c|c|c|}
\multicolumn{6}{c}{Wide lens} \\
\hline
\multicolumn{3}{c}{$F_{\rm UL}$} & \multicolumn{3}{c}{$F_{\rm DL}$} \\
\hline
BW$_{\rm -3dB}$ & FBR & R$_{\rm rad}$ & BW$_{\rm -3dB}$ & FBR & R$_{\rm rad}$ \\
deg. & dB & $\Omega$ & deg. & dB & $\Omega$ \\
\hline
66.5 & -30.2 & 63 & 64 & -29.6 & 115 \\
\hline
\multicolumn{6}{c}{Narrow lens} \\
\hline
\multicolumn{3}{c}{$F_{\rm UL}$} & \multicolumn{3}{c}{$F_{\rm DL}$} \\
\hline
BW$_{\rm -3dB}$ & FBR & R$_{\rm rad}$ & BW$_{\rm -3dB}$ & FBR & R$_{\rm rad}$ \\
deg. & dB & $\Omega$ & deg. & dB & $\Omega$ \\
\hline
82 & -18.5 & 64 & 85 & -20.1 & 90 \\
\hline
\end{tabular}
\end{table}

\subsection{Selected observations on the antenna behavior}

In this section we have gathered some typical (qualitative)
tendencies in the antenna behavior noticed during the the simulation
process. One of the most important observations with the simulated
structures (having a finite size reflector or no reflector at all)
is the fact that a good FBR level contradicts rather heavily with a
narrow beam width in H-plane. In other words, with nearly all the
simulated lens topologies and reflector shapes, narrowing the
H-plane beam width deteriorates the FBR (this can not readily be
seen in the results presented for the prototype with two modes,
since the whole lens topology is changed to manipulate the H-plane
beam width). With a fixed topology of the lens (e.g. the wide lens
topology in the prototype), the most effective way to narrow the
H-plane beam width is to increase the value of the lattice parameter
$b$, see Fig.~\ref{proto_geom} for definition. This, however, has
also a strong negative effect on the FBR level. If a rather moderate
H-plane beam width is desired, the rows of wires can be brought in a
close vicinity of each other still maintaining the FBR at a very
good level \cite{Ikonen}. If a considerably narrow H-plane beam
width is to be achieved, the wave has to travel a longer distance in
the lens (phase shifts of the currents induced to the wires far away
from the radiator have to change rapidly enough). This means
increasing the depth of the lens.

A general tendency with the simulated structures is that a good FBR
level corresponds to a small value of the radiation resistance and
vice versa. One of the biggest drawbacks of the studied structures
is the strong deviation of the input impedance (radiation
resistance) over the desired frequency range. Typically the change
in the radiation resistance over the UMTS band is close to 50
percent, the minimum resistance occurring at lower frequencies. This
naturally implies difficulties in maintaining the matching at a
reasonable level over the whole frequency range, but also causes
deviation in the beam width and FBR. With most of the studied
structures the beam width is approximately 5$-$10$^{\circ}$ wider at
the UL range than at the DL range (the aperture is narrower in terms
of wavelength) and the deviation in the FBR is approximately 5 dB.

The value of the loading capacitance naturally depends  on the
insertion period of the loads. However, a suitable capacitance range
producing reasonable performance can be found and the following
observations can be made for this range (a fixed lens topology close
to a triangular one assumed): Increasing the capacitance narrows the
beam width but also deteriorates the FBR. This is due to the fact
that with increasing capacitance the radiation resistance grows (up
to a certain limit). When increasing the capacitance too much the
radiation pattern changes very dramatically leading to a pattern
with a very strong backward lobe and the main beam of a shape of a
trefoil. On the other hand, a too low value for the capacitance
leads to a structure that practically does not radiate (due to the
destructive  interaction between the source and the passive wires).

\section{FDTD model for the dispersive dielectric lens}

\begin{figure*}[b!]
\centerline{\subfigure[The exact simulated structures:  Left,
simulated structure in FEKO (3D structure), right, simulated
structure in FDTD (2D
structure).]{\includegraphics[width=8.8cm]{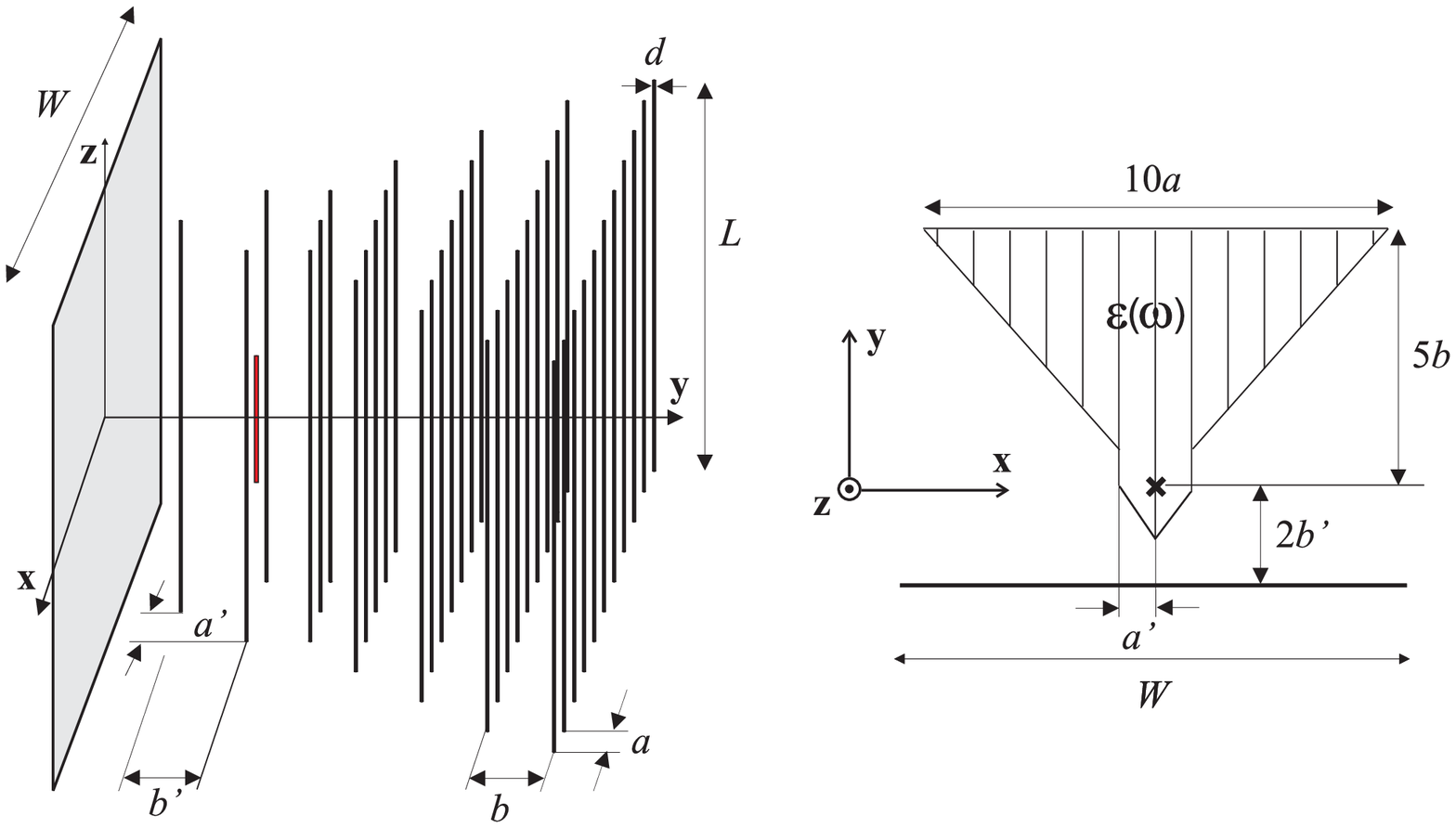}
\label{FDTDsche}} \hfill \subfigure[Comparison between the simulated
results.]{\includegraphics[width=8.0cm]{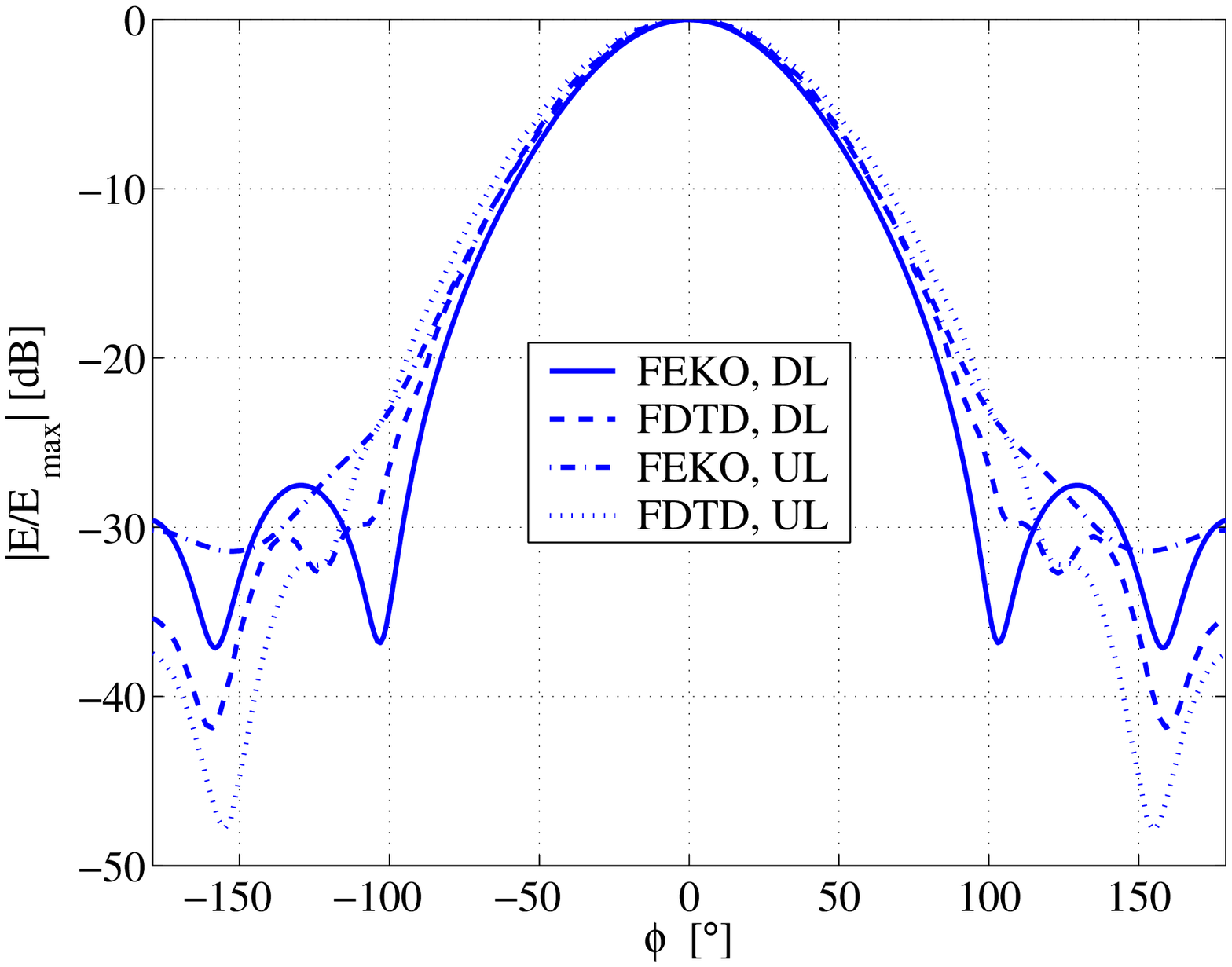}
\label{FDTDres}}} \caption{(a) Simulated structures. (b) Comparison
between the results given by FEKO and FDTD.} \label{FDTD_res}
\end{figure*}

In this section we consider a finite size lattice of loaded wires as
a block of dispersive dielectric material, as schematically depicted
in Fig.~\ref{lens}. The dispersive behavior of the capacitively
loaded wire medium has been shown to follow the expression
(\ref{permittivity}) \cite{Belov-PhysE} (when the operational
frequency is far below the first stop band and the array is dense).
In this section we use an in-house finite-difference time-domain
(FDTD) code and study the equivalence of the wire lattice to a
dielectric lens in the case of a \emph{moderate number of lattice
periods}.

We choose to analyze the wide lens (for the transversal geometry see
Fig.~\ref{fig_first_case}). It was experimentally shown in
\cite{Ikonen} that the height of the structure (the dimension in the
$z$-direction) has only a very weak effect on the radiation pattern
in the H-plane at low frequencies. Therefore to reduce the
computational burden we consider without a significant loss of
accuracy a two dimensional case in FDTD. The exact simulated
structures are depicted in Fig. \ref{FDTDsche}.

The dispersive behavior for the effective permittivity  has been
implemented in the code using the auxiliary differential equation
(ADE) method \cite{Taflove}. The ADE method is applied for a medium
having Lorentzian type dispersive permittivity with one pair of
poles in its susceptibility response. For this kind of medium
Ampere's law can be expressed in time domain as \e \nabla \times
\textbf{H}(t) = \E_0\E_{\infty}\frac{d}{dt}\textbf{E}(t) +
\sigma{\textbf{E}(t)} + \textbf{J}_0(t), \label{Ampere} \f where
$\E_{\infty}$ is the value of the relative permittivity at infinite
frequency, $\sigma$ is the conductivity of metal parts in the medium
and $\textbf{J}_0(t)$ is the polarization current associated with
the zeroth Lorentz pole pair. The evident generalization of
(\ref{permittivity}) for the case of lossy media is \e \E_{\rm
eff}(\o) = \E_0\bigg{(}1 + \frac{C/(\E_0s^2)} {1
+2j\o\delta_o/\o_0^2 - \o^2/\o_0^2}\bigg{)} = \E_{\infty} +
\frac{(\E_{\rm s} - \E_{\infty})\omega_0^2}{\omega_0^2 +
2j\omega\delta_0 - \omega^2}, \label{eps_ADE} \f where $\E_{\rm s}$
is the static permittivity, $\omega_0$ is the undamped frequency of
the zeroth pole pair (the undamped resonant frequency of the wire
medium) and $\delta_0$ is the damping factor associated with
$\omega_0$ (in the effective medium model (\ref{permittivity})
$\delta_0=0$). Utilizing (\ref{Ampere}) and (\ref{eps_ADE}) we
construct a three step fully explicit procedure for updating the
field components and the components of the polarization current.

Fig.~\ref{FDTDres} shows the normalized far field  patterns for the
electric field simulated with \verb"FEKO" and FDTD. This figure
demonstrates   the equivalence of the wire lattice to a finite size
dielectric block. The agreement between the shape of the main lobe
is good at both frequencies. The overall agreement of the two
patterns is good at the DL frequency. When thinking of dielectric
lenses implemented with substrates, the wire medium lens behaves in
a similar manner, however, it is cheaper and, more importantly,
\emph{lighter}.

\section{Prototype antenna}

To manufacture the prototype we use a standard printed circuit board
technique where the loaded wires are split strips on dielectric
plates.  The effects of the dielectric sheets have been taken into
account by slightly modifying the prototype design, see Table 2. The
modifications were enough to suppress the FBR with the cost of
increased reflector size. Note, however, that the use of a larger
size reflector is well justified since when mounting the antenna
structure into a base station mast the mast itself operates as a
large reflector.

\begin{table}[b!]
\begin{center}
\renewcommand{\arraystretch}{1.3}
\caption{The parameters of the implemented prototype (see
Fig.~\ref{proto_geom} for the parameter definition).}
\begin{tabular}{|c|c|c|c|c|c|c|c|c|}
\hline
$a'$ & $b'=b$ & $a$ & $d$ & $C_{\rm f}$ & $C_{\rm w}$ & $C_{\rm n}$ & $L$ & $W$ \\
mm & mm & mm & mm & pF & pF & pF & mm & mm\\
\hline
9.0 & 12.5 & 11.5 & 1.125 & 0.125 & 0.10 & 0.13 & 200 & 200 \\
\hline
\end{tabular}
\smallskip

$\dag$ {\small now $a$ and $b$ are the same for both lenses.}
\end{center}
\label{proto_param}
\end{table}

A photo showing the implemented prototype (with the wide lens
connected) is presented in Fig.~\ref{protop}. FR-4 is used as the
dielectric substrate material in the prototype antenna (the
substrate thickness equals $1.0$ mm with the wide lens and $0.8$ mm
with the narrow lens, $\epsilon_r\approx4.5$). The strip width is
equal to $t_{\rm w}=2.25$ mm with the wide lens and $t_{\rm n} =
2.50$ mm with the narrow lens. The period of the gaps (the insertion
period of the loads) is equal to $l=5.55$ mm. To choose the gap
width $w$, the following approximate expression was used: \e C_0= m
\times \epsilon_0{t(\epsilon_r+1)\over \pi}\log{8l\over w}, \f where
the empirically found multiplication factor $m$ is approximately
0.7. One half-wavelength dipole was used as an active source. The
dipole was fed with a coaxial cable having $\lambda/4$ coaxial balun
symmetrizing the feeding.

The measurements validating the main practical importance of the
present work, the feasibility of the wire medium lens as a beam
shaping element, are the measurement of the $S_{\rm 11}$-parameter
and the measurement of the 2D radiation pattern in the H-plane.
These measurements confirm that the antenna matching does not change
significantly when switching from one beam width to another, and
that the beam width can be controlled by manipulating the lattice
topology.

\begin{figure*}[t!]
\centerline{\subfigure[Prototype photo.]
{\includegraphics[width=5.5cm]{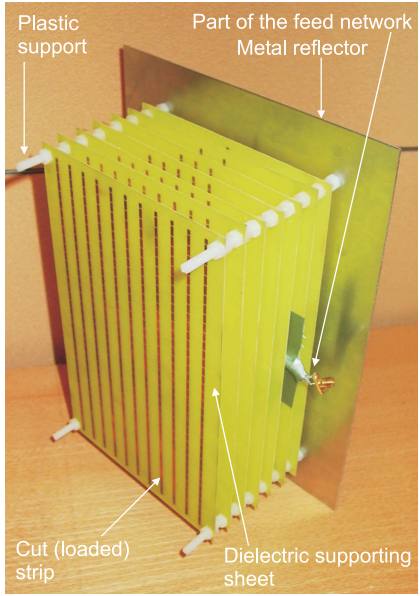} \label{protop}}
\hfil \subfigure[The measured matching level: Solid line, wide lens,
dashed line, narrow
lens.]{\includegraphics[width=8.0cm]{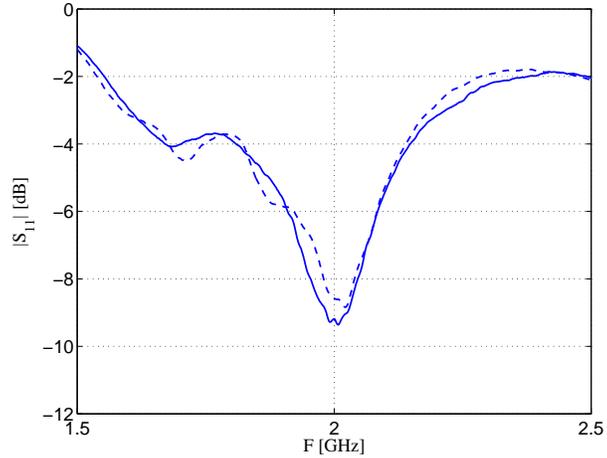}
\label{matching}}} \caption{(a) A photograph showing the implemented
prototype with the wide lens. (b) The measured matching level.}
\label{proto_photo}
\end{figure*}

\subsection{Matching level}

Fig.~\ref{matching} shows the measured $S_{\rm 11}$-parameter. The
matching of the antenna has not been optimized (the best matching
occurs a bit below the center frequency 2.045 GHz). A matching
network (implemented e.g. using coaxial stubs) would improve the
matching level at the center frequency and offer a possibility to
tune the location of the $S_{\rm 11}$-minimum. However, when
thinking of a real end product, special attention should be devoted
to the design of a wideband matching network.

With the introduced prototype the matching level remains almost the
same for both wide and narrow lens. However, from the presented
measurement we can not readily draw a general conclusion that the
matching hardly depends on the removable part of the dielectric
lens. This is due to the fact that the prototype antenna is only
satisfactorily matched, with a well matched antenna the effect seen
on the matching level can be larger.

\subsection{2D radiation pattern}

Fig.~\ref{meas_res} introduces the measured H-plane radiation
patterns for both wide and narrow lenses. The \verb"FEKO"
simulations for the implemented prototype are plotted for reference.
The measured half-power beam widths and FBRs are gathered in Table
\ref{measres}. The measured gain values (at the center frequency
$F_{\rm c}$= 2.045 GHz) with the wide and narrow lens are 9.8 and
8.0 dBi, respectively.

The measured results agree rather well with the simulated ones. The
main lobe beam width and the FBR are in good agreement. The
deviation is visible only in the side lobe directions. This
deviation and the slight asymmetry of the pattern are, probably,
caused by the feed  network and the currents induced to the edges of
the reflector plate. The reflectivity level of the measurement
chamber was estimated to be $-$25 dB at 2 GHz (implying that the
side and back lobe levels are in fact unreliable). The manufacturing
process is fast, robust and inexpensive. However, a random
discrepancy of approximately $\pm$10 percents in the load
capacitance was measured over the slots with a high precision
$LC-$meter.

\begin{figure*}[b!]
\centerline{\subfigure[Wide lens: ``UL'' corresponds to  $F = F_{\rm
UL} = 1.95$ GHz, ``DL'' corresponds to $F = F_{\rm DL} = 2.14$
GHz.]{\includegraphics[width=8.0cm]{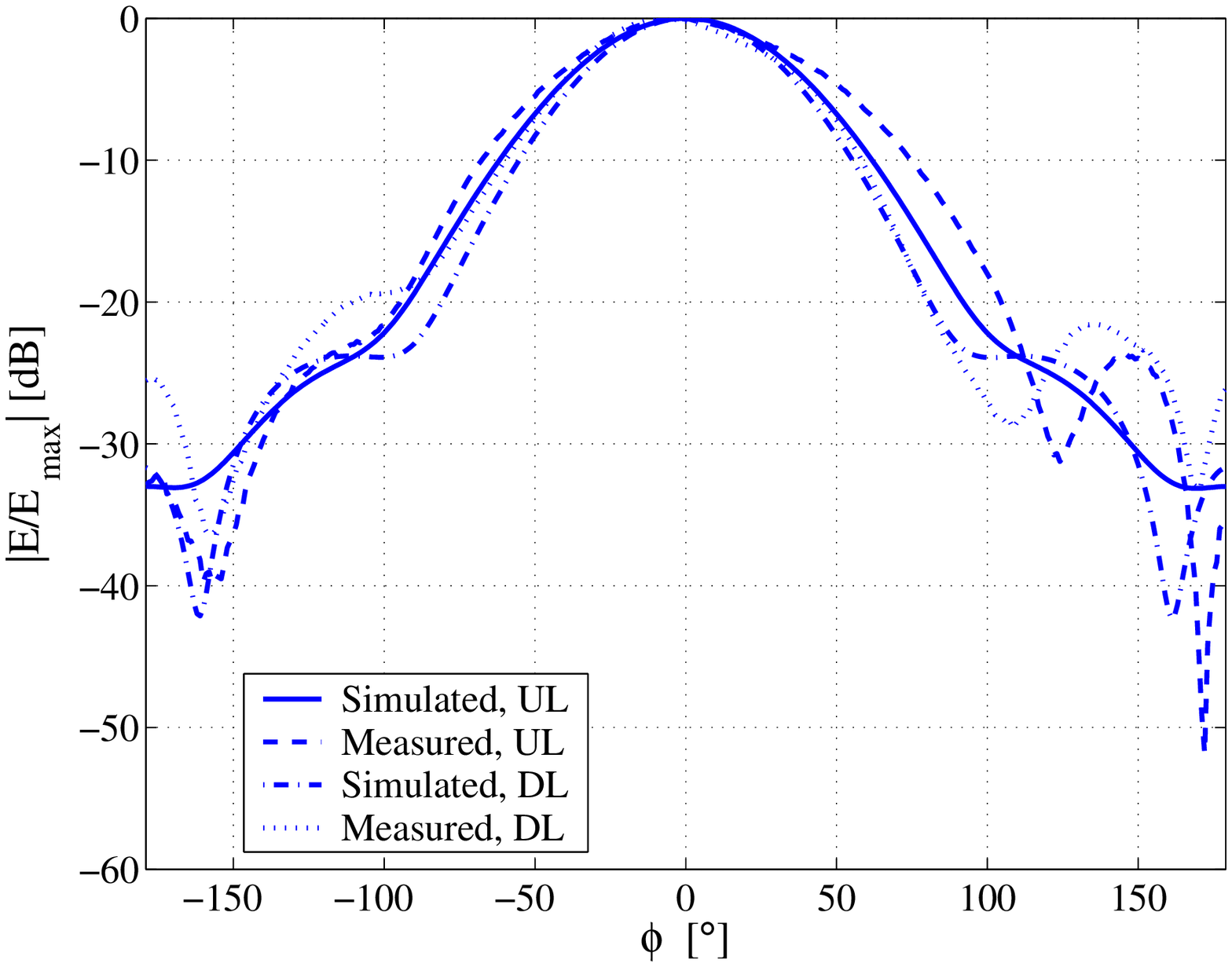} \label{wlm}}
\hfil \subfigure[Narrow lens: ``UL'' corresponds to $F = F_{\rm UL}
= 1.95$ GHz, ``DL'' corresponds to $F = F_{\rm DL} = 2.14$
GHz.]{\includegraphics[width=8.0cm]{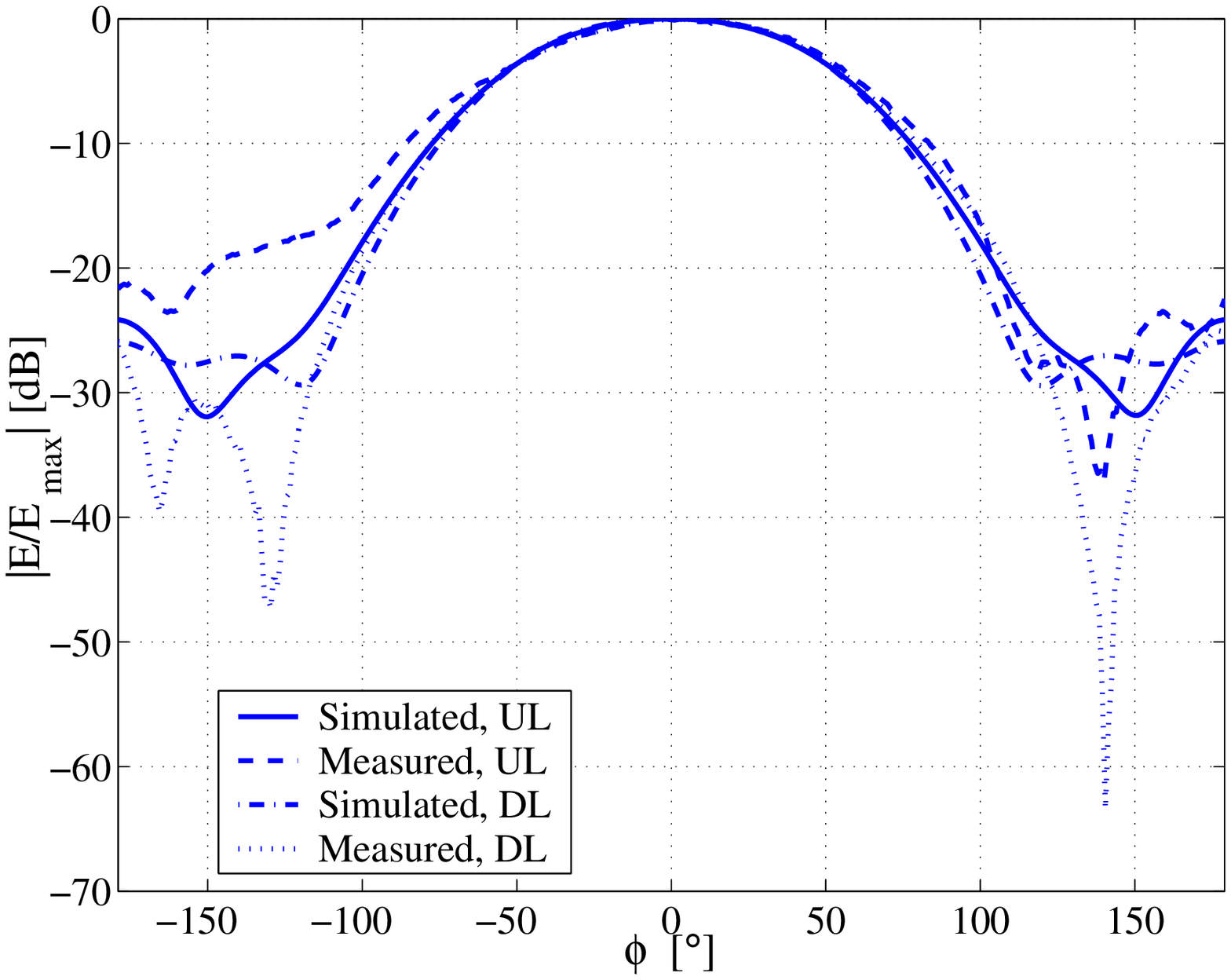} \label{nlm}}}
\caption{The measured and simulated H-plane radiation patterns.}
\label{meas_res}
\end{figure*}

\begin{table}
\renewcommand{\arraystretch}{1.3}
\caption{The measured radiation characteristics.} \label{measres}
\centering
\begin{tabular}{|c|c||c|c|}
\multicolumn{4}{c}{Wide lens} \\
\hline
\multicolumn{2}{c}{$F_{\rm UL}$} & \multicolumn{2}{c}{$F_{\rm DL}$} \\
\hline
BW$_{\rm -3dB}$  & FBR & BW$_{\rm -3dB}$ & FBR \\
deg. & dB & deg. & dB \\
\hline
74 (66) & -33 (-33) & 62 (60) & -25.5 (-32) \\
\hline
\multicolumn{4}{c}{Narrow lens} \\
\hline
\multicolumn{2}{c}{$F_{\rm UL}$} & \multicolumn{2}{c}{$F_{\rm DL}$} \\
\hline
BW$_{\rm -3dB}$  & FBR & BW$_{\rm -3dB}$ & FBR \\
deg. & dB & deg. & dB \\
\hline
90 (91) & -24 (-24) & 88 (92) & -26 (-26) \\
\hline
\end{tabular}
\smallskip

$^\dag$ {\small the value in the brackets is the simulated result.}
\end{table}

\section{Conclusion}

In the present paper we have studied the applicability  of the
loaded wire medium as a beam shaping lens. The dispersion properties
of the loaded wire medium introduced in the literature have been
briefly revised. Contrary to the known EBG beam shaping techniques,
we have chosen to utilize the frequency range far below the first
stop band of the structure. In this region the loaded wire medium
can be interpreted as a continuous artificial dielectric. A concept
for a compact dual-mode base station antenna operating in the UMTS
frequency range has been presented. We have confirmed that the
theory of an aperture radiator is applicable when designing the wire
medium lens antenna. A FDTD model for the dispersive dielectric lens
having the wire lattice geometry has been constructed. The full wave
simulations reveal that the loaded wire lattice operates as a
dielectric block even with a rather small number of lattice periods.
A promising performance of the prototype is shown to be achieved
with a rather simple structure and cheap manufacturing process.

\section*{Acknowledgement}

This work has been supported in part by Nokia Research Center,
Filtronic LK, and TEKES.


\newpage

\begin{thebibliography}{99}

\bibitem{Webb}
W. Webb, {\it The future of wireless communications},  Boston:
Artech House, 2001.

\bibitem{Mishra}
A. R. Mishra, {\it Fundamentals of cellular network  planning and
optimization}, New York: John Wiley $\&$ Sons, 2004.

\bibitem{Brown}
J. Brown, ``Artificial dielectrics'',  {\it Progress in
dielectrics}, vol.~2, pp.~195--225, 1960.

\bibitem{Rotman}
W. Rotman, ``Plasma simulation by artificial  and parallel plate
media,'' {\it IRE Trans. Ant. Propagat.}, vol.~10, pp.~82--95, 1962.

\bibitem{Joannopoulos}
J. D. Joannopoulos, R. D. Mead, and J. N. Winn  {\it Photonic
crystals: Molding the flow of light}, Princeton University Press,
Princeton, NJ, 1995.

\bibitem{Kuzmiak}
V. Kuzmiak, A. A. Maradudin, and F. Pincemin, ``Photonic band
structures of two-dimentional systems containing metallic
components,'' {\it Phys. Rev. B}, vol.~50, no.~23, pp.~16835--16844,
1994.

\bibitem{Nicorovici}
N. A. Nicorovici, R. C. McPhedran, and L. C. Botten ``Photonic band
gaps for arrays of perfectly conducting cylinders,'' {\it Phys. Rev.
E}, vol.~52, no.~1, pp.~1135--1145, 1995.

\bibitem{Sigalas}
M. M. Sigalas, C. T. Chan, K. M. Ho, and C. M. Soukoulis,
``Metallic photonic band-gap materials,'' {\it Phys. Rev. B},
vol.~52, no.~16, pp.~11744--11751, 1995.

\bibitem{Pitarke}
J. M. Pitarke, F. J. Garc\'ia-Vidal, and J. B. Pendry,  ``Effective
electronic response of a system of metallic cylinders,'' {\it Phys.
Rev. B}, vol.~57, no.~24, pp.~15261–-15266, 1998.

\bibitem{Moses-Engheta}
C. A. Moses and N. Engheta,  ``Electromagnetic wave propagation in
the wire medium: A complex medium with long thin inclusions,'' {\it
Wave Motion}, vol.~34, pp.~301--317, 2001.

\bibitem{Belov-JEWA}
P. A. Belov and S. A. Tretyakov, ``Dispersion and reflection
properties of artificial media formed by regular lattices of ideally
conducting wires'', {\it J. Electromagnetic Waves and Applications},
vol.~16, no.~8, pp.~1153--1170, 2002.

\bibitem{Belov-PhysE}
P. A. Belov, C. R. Simovski, and S. A. Tretyakov, ``Two-dimensional
electromagnetic crystals formed by reactively loaded wires,'' {\it
Phys. Rev. E}, vol.~66, 036610, 2002.

\bibitem{Belov-PhysB}
P. A. Belov, R. Marqu\'es, S. I. Maslovski, I. S. Nefedov, M.
Silveirinha , C. R. Simovski, and S. A. Tretyakov, ``Strong spatial
dispersion in wire media in the very large wavelength limit'', {\it
Phys. Rev. B}, vol.~ 67, 113103, 2003.

\bibitem{Maslovski}
S. I. Maslovski, S. A. Tretyakov, and P. A. Belov,  ``Wire media
with negative effective permittivity: A quasi-static model,'' {\it
Microwave Opt. Technol. Lett.}, vol.~35, no.~1, pp.~47--51, 2002.

\bibitem{Lourtioz}
J.--M. Lourtioz, A. de Lustrac, F. Gadot, S. Rowson, A. Chelnokov,
T. Brillat, A. Ammouche, J. Danglot,  O. Vanb\'{e}sien, and D.
Lippens ``Toward controllable photonic crystals for centimeter- and
millimeter-wave devices,'' {\it J. Lightwave Technol.}, vol.~17,
no.~11, pp.~2025--2031, 1999.

\bibitem{Lustrac1}
A. de Lustrac, F. Gadot, E. Akmansoy, and T. Brillat,
``High-directivity planar antenna using controllable photonic
bandgap material at microwave frequencies,'' {\it Appl. Phys.
Lett.}, vol.~78, no.~26, pp.~4196--4198, 2001.

\bibitem{Poilasne_the}
G. Poilasne, J. Lenormand, P. Pouliguen,  K. Mahdjoubi, C. Terret,
and Ph. Gelin, ``Theoretical study of interactions between antennas
and metallic photonic bandgap materials,'' {\it Microwave Opt.
Technol. Lett.}, vol.~15, no.~6, pp.~384--389, 1997.

\bibitem{Poilasne2}
G. Poilasne, P. Pouligen, K. Mahdjoubi, C. Terret,  Ph. Gelin, and
L. Desclos, ``Experimental radiation pattern of dipole inside
metallic photonic bandgap material,'' {\it Microwave Opt. Technol.
Lett.}, vol.~22, no.~1, pp.~10--16, 1999.

\bibitem{Milne}
R. Milne, ``Dipole array lens antenna,''  {\it IEEE Trans.\ Antennas
Propagat.,} vol.~AP--30, no.~4, pp.~704--712, 1982.

\bibitem{Silveirinha2}
M. G. M. V. Silveirinha, C. A. Fernandes,  ``Design of a
non-homogeneous wire media lens using genetic algorithms'', {\it
Antennas and Propagation Society International Symposium}, Columbus,
Ohio USA, June 22--27, pp.~730--733, 2002.

\bibitem{Simovski1}
C. R. Simovski and S. He,  ``Antennas based on modified metallic
photonic bandgap structures consisting of capacitively loaded
wires,'' {\it Microwave Opt. Technol. Lett.}, vol.~31, no.~3,
pp.~214--221, 2001.

\bibitem{Ikonen}
P. Ikonen, C. Simovski, and S. Tretyakov,  ``Compact directive
antennas with a wire-medium artificial lens,'' {\it Microwave Opt.
Technol. Lett.}, vol.~43, no.~6, pp.~467--469, 2004.

\bibitem{Tayeb}
G. Tayeb and D. Maystre, ``Rigorous theoretical study of finite-size
two-dimensional photonic crystals doped by microcavities,'' {\it J.
Optical Soc. America A}, vol.~14, no.~12, pp.~3323--3332, 1997.

\bibitem{Cheype}
C. Cheype, C. Serier, M. Th\'evenot, T. Mon\'edi\'ere, A. Reineix,
and B. Jecko, ``An electromagnetic bandgap resonator antenna,'' {\it
IEEE Trans. Antennas Propagat.}, vol.~50, no.~9, pp.~1285--1290,
2002.

\bibitem{Temelkuran}
B. Temelkuran, M. Bayindir, E. Ozbay, R. Biswas, M. M. Sigalas, G.
Tuttle, and K. M. Ho,  ``Photonic crystal-based resonant antenna
with a very high directivity,'' {\it J. Appl. Phys.}, vol.~87,
no.~1, pp.~603--605, 2000.

\bibitem{Kathrein}
The www--page of KATHREIN-Werke\\
http://www.kathrein.de


\bibitem{Feko}
The www--page of FEKO\\http://www.feko.co.za/

\bibitem{Balanis}
C. A. Balanis, {\it Antenna theory: Analysis and design},  New York:
John Wiley, 1997.

\bibitem{Taflove}
A. Taflove, {\it Computational electrodynamics: The
finite-difference time-domain method}, Artech House, 1995.

\end{thebibliography}
\end{document}